\documentclass[runningheads]{article}
\usepackage{graphicx}
\begin{document}
%
\title{Fine tuning U-Net for ultrasound image segmentation: which layers?}
%
%
\author{Mina Amiri \and
Rupert Brooks \and
Hassan Rivaz}
%
%
%
\maketitle              
\begin{abstract}
Fine-tuning a network which has been trained on a large dataset is an alternative to full training in order to overcome the problem of scarce and expensive data in medical applications. While the shallow layers of the network are usually kept unchanged, deeper layers are modified according to the new dataset. This approach may not work for ultrasound images due to their drastically different appearance. In this study, we investigated the effect of fine-tuning different layers of a U-Net which was trained on segmentation of natural images in breast ultrasound image segmentation. Tuning the contracting part and fixing the expanding part resulted in substantially better results compared to fixing the contracting part and tuning the expanding part. Furthermore, we showed that starting to fine-tune the U-Net from the shallow layers and gradually including more layers will lead to a better performance compared to fine-tuning the network from the deep layers moving back to shallow layers. We did not observe the same results on segmentation of X-ray images, which have different salient features compared to ultrasound, it may therefore be more appropriate to fine-tune the shallow layers rather than deep layers. Shallow layers learn lower level features (including speckle pattern, and probably the noise and artifact properties) which are critical in automatic segmentation in this modality.

keywords: Ultrasound imaging, Segmentation, Transfer learning, U-Net.
\end{abstract}
\section{Introduction}

Training a deep convolutional neural network (CNN) from scratch is challenging, especially in medical applications, where annotated data is scarce and expensive. An alternative to full training is transfer learning, where a network which has been trained on a large dataset is fine-tuned for another application. When the new dataset is small, the recommended approach in fine-tuning is to keep the first layers of the network unchanged, and to fine-tune the last layers \cite{Yosinski2014}. It is shown that first layers of a CNN represent more low-level features, while more semantic and high-level features are recognized by deeper layers \cite{LeCun2015}. Therefore, fine-tuning the deepest layers originates from the assumption that basic features of the datasets (associated with shallow layers) are similar, and more specific features of the datasets (associated with deeper layers) should be tuned in order to get acceptable results in a different application. This assumption may not hold true in some medical applications. For instance, in ultrasound imaging the presence of wave-tissue interactions such as scattering lead to creation of speckles, which may not be present in natural images or images from other medical modalities.

Ultrasound imaging is a standard modality for many diagnostic and monitoring purposes, and there has been significant research into developing automatic methods for segmentation of ultrasound images \cite{Looney2018,Yang2017}. U-Net \cite{Ronneberger2015} for instance has been shown to be a fast and precise solution for medical image segmentation, and has successfully been adapted to segment ultrasound images too \cite{Yang2019,Alsinan2019,Wang2018,Yap20182}. In this study, we investigate the effect of fine-tuning different layers of a U-Net network for the application of ultrasound image segmentation. We hypothesize that ultrasound-specific patterns are learned in shallow layers which disentangle the information in speckle pattern. Therefore, fine-tuning these layers is critical in fine-tuning the weights learned from another domain.

\section{Methodology}
This section provides an overview of the datasets used in this study, details of pre training and fine-tuning the U-Net, and the performance metrics used to validate our results.
\subsection{Datasets}
\begin{figure}
\hspace{-2cm} \includegraphics[scale=1.01]{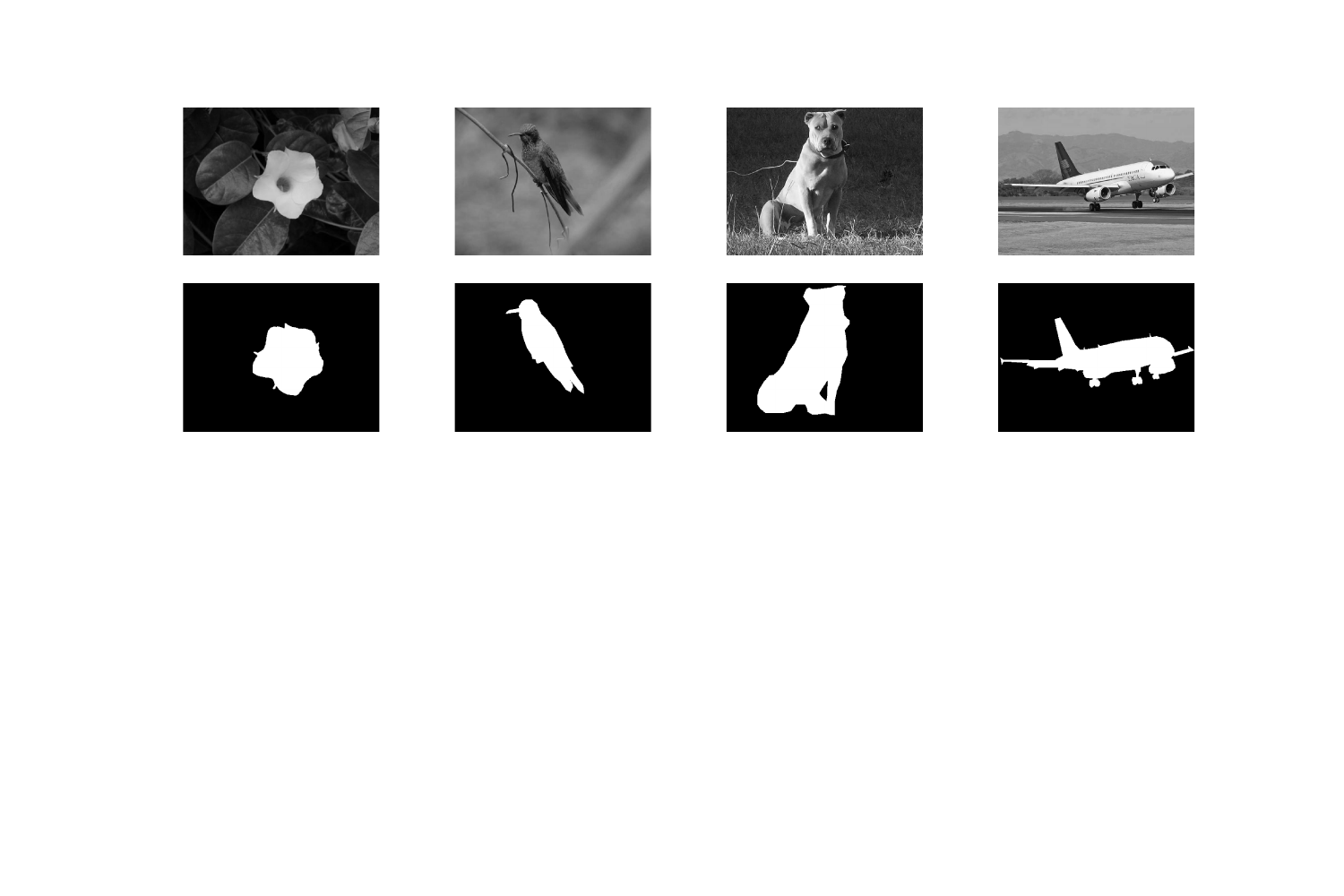}
\vspace{-5.5cm}
\caption{Some examples from the XPIE dataset and the associated masks. These images have very different appearances when compared to X-ray or ultrasound images.} \label{fig1}
\end{figure}

In order to pre-train the network, we used the XPIE dataset which contains 10000 segmented natural images \cite{Xia2017}. The images in this dataset are not gray scale. In order to have a more similar pre-training dataset to ultrasound dataset, we converted these images into black and white prior to feeding to the network. We used 40 epochs to train the network, and 10\% of the data was considered as the validation set. Figure \ref{fig1} shows a few examples of this dataset. The pre-trained network was then used for the task of segmentation of ultrasound B-mode images. The ultrasound imaging dataset contains 163 images of the breast with either benign lesions or malignant tumors \cite{Yap2018}. In order to investigate whether the results are specific to the ultrasound imaging, we repeated the analysis for a chest X-ray dataset with the total of 240 images \cite{Vanginneken2006}, wherein we used the pre-trained network to segment both lungs.
\subsubsection{Data Augmentation} As the size of ultrasound and X-ray datasets was small, we implemented data augmentation techniques to improve the network performance, invariance and robustness. For these datasets, the network should be robust to shift, rotation, flipping, shearing and zooming. We generated smooth deformations of images using random and small degrees of all these transformations. In total, we had 600 images including the original images to train the network. In the case of natural images, we did not augment the data.

\subsection{Analysis}
We used the same U-Net architecture introduced in the original paper \cite{Ronneberger2015} except that we used up-sampling in the expanding path. The network consists of blocks of two convolutional layers with ReLU activation, followed by either a maxpooling or an upsampling operation. There are 64 filters in both layers in the first block. Following each maxpooling operation, the number of filters doubles, while after each upsampling operation, the number of filters is halved. A 1 x 1 convolutional layer with sigmoid activation is used as the last layer to map the feature vector to the interval of 0 and 1. For evaluation purposes, pixels with the value above 0.5 were considered as 1, and pixels with the value below 0.5 were considered as 0. We did not use batch normalization, but we used the dropout technique after the contracting path.  

We first trained a U-Net using the XPIE dataset. The parameters of this pre-trained network was then utilized as an initial point to retrain the network for ultrasound or X-ray image segmentation. All images were resized to 256 x 256 pixels and were normalized to [0,1]. To examine whether fine-tuning shallow or deep layers differ significantly, we divided the U-Net into two parts: contracting (up to the 10th convolutional layer) and expanding (from 10th convolutional layer to the end). While freezing one part, we fine-tuned the other part using the ultrasound B-mode images as the training data. We then switched the frozen and trainable parts.

\begin{figure}
\centering
\includegraphics[ scale=.3]{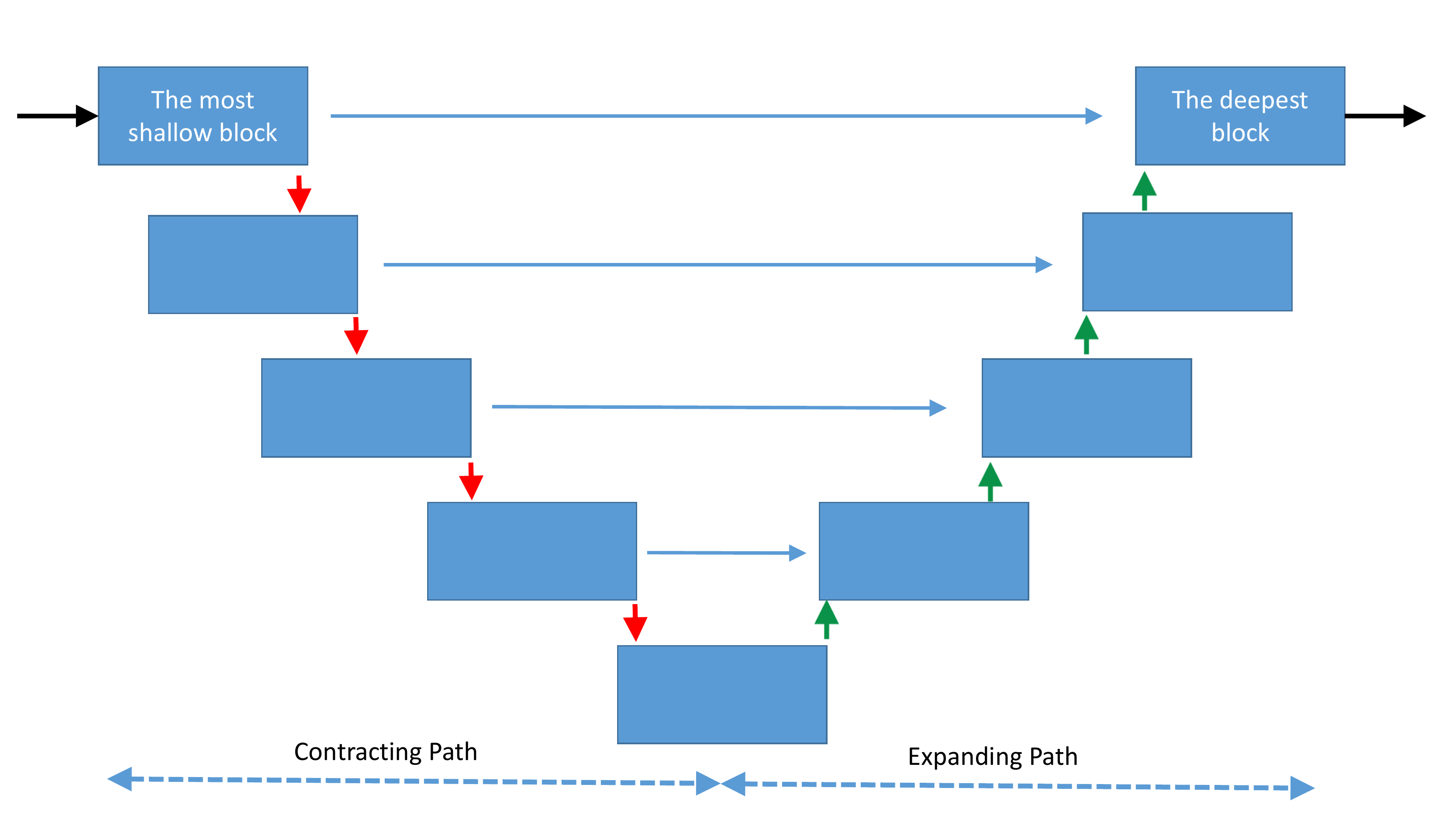}
\caption{Schematic of U-Net. Each box represents one block of layers. Red, green and blue arrows respectively represent maxpooling, upsampling and copy-crop-concatenating.} \label{fig2}
\end{figure}

Next, we repeated the same approach but in a finer manner. We grouped all layers between two consecutive maxpooling or up-sampling layers in to one block (Fig \ref{fig2}). Each block therefore consisted of two convolutional layers. We started by fine-tuning the first block (first two layers) while freezing all other layers. We then included other blocks in the fine-tuning procedure one-by-one, until the whole network was trained (from shallow to deep layers). We repeated the same procedure in the opposite direction; we started fine-tuning the deepest block while freezing the remaining of the network, and then included more blocks in fine-tuning until the whole network was trained (from deep to shallow layers). The same analysis was done for the chest X-ray dataset to segment the lungs.

\subsection{Performance Metrics}
We used 5-fold cross validation to evaluate the performance of the network. All the data was randomly divided into five folds, with four folds used for training and the 5th fold used for validation. The procedure was repeated five times so that all five folds served as the validation set. To evaluate the performance of the network in segmenting the images, we used Dice score, pixel error and rand score. Dice score equals twice the number of elements correctly predicted as the mask (2 * TP) divided by the sum of the number of elements in the ground-truth mask (TP + FN) and the predicted mask (TP + FP). Pixel error is the percentage of voxels falsely predicted by the network, and rand score is a measure of similarity between two clusterings by considering pairs that are assigned in the same or different clusters in the predicted and ground-truth image \cite{Rand1971}. In this study, we used the rand score adjusted for chance clustering.  

\section{Results}
\begin{figure}
\hspace{-1.8cm}\includegraphics[scale=1.01]{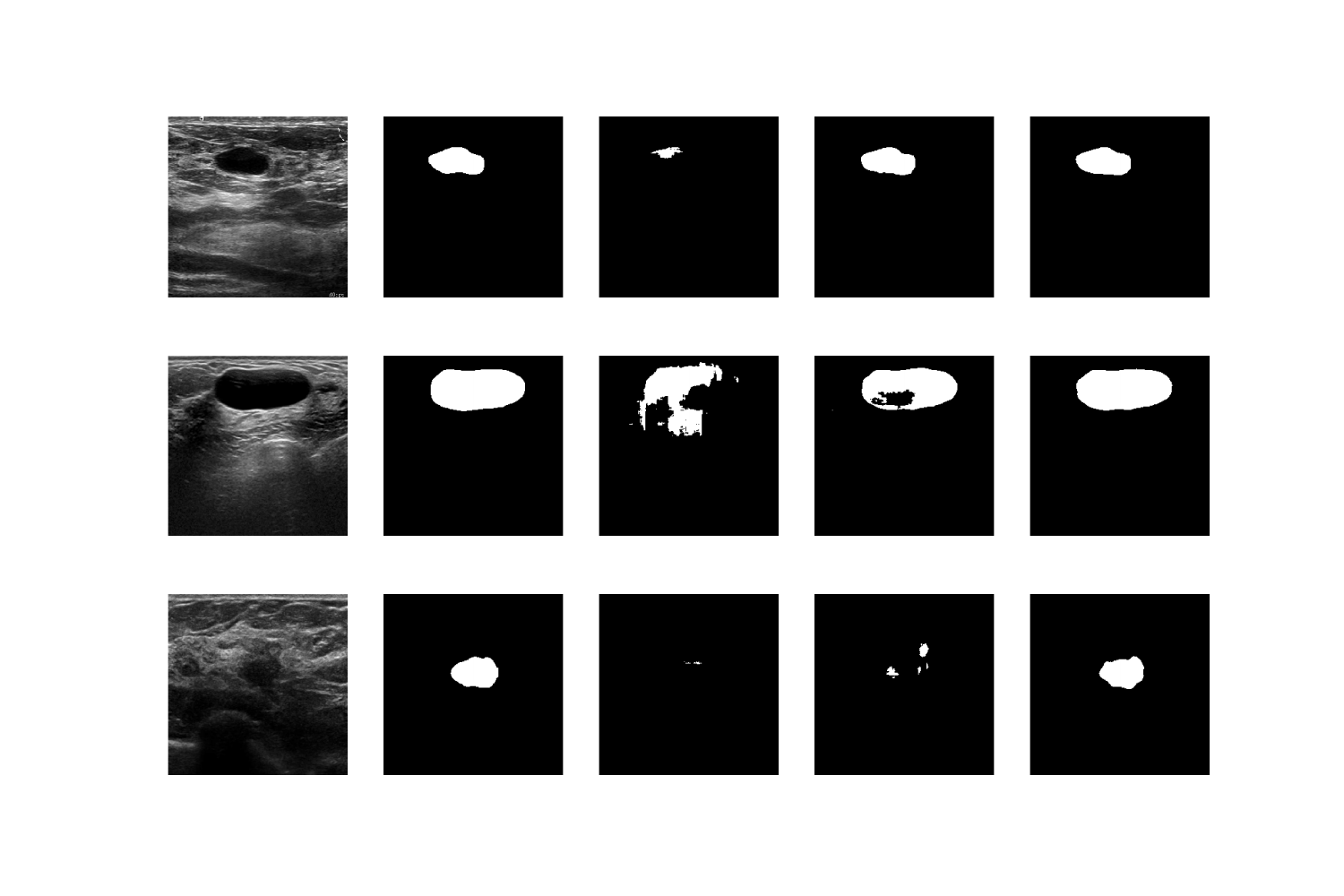}
\vspace{-1.5cm}
\caption{Comparison of the two scenarios on a few examples. From left to right: the original image, the ground truth mask, the predicted mask by the pre-trained network, the predicted mask when the expanding part is fine-tuned, the predicted mask when the contracting part is fine-tuned. First row: both scenarios work well. Second row: fine-tuning the contracting path outperforms the other scenario. Third row: fine-tuning the contracting path performs much better than the other scenario. } \label{fig3}
\end{figure}

\begin{figure}
\includegraphics[width=\textwidth]{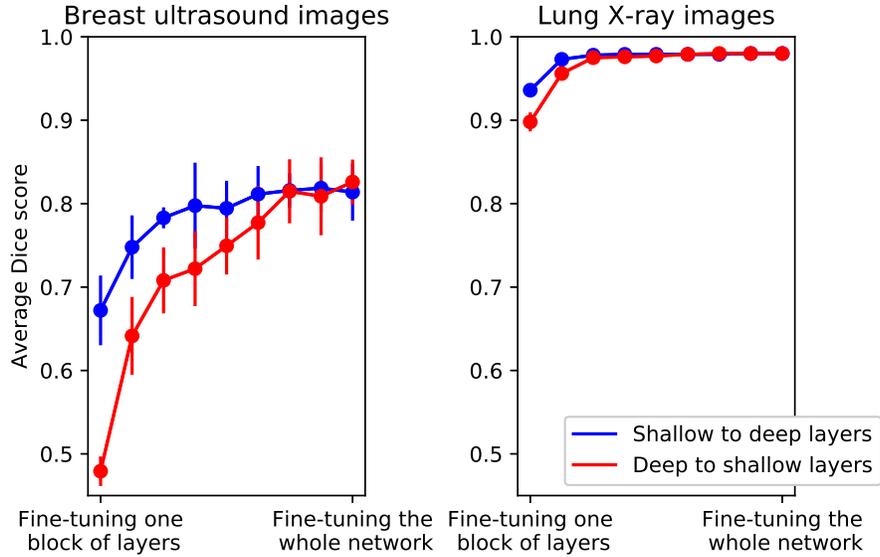}
\hspace{1.5cm}
\caption{Average Dice score for two different scenarios for ultrasound and X-ray images. Error bars depict the standard deviation of the mean among the five folds.} \label{fig4}
\end{figure}

Training the contracting part and freezing the expanding part led to better results compared to freezing the contracting part and fine-tuning the expanding part (Dice score: 0.80 $\pm$ 0.03 vs. 0.72 $\pm$ 0.04, pixel error: 1.4\% $\pm$ 0.5 vs. 1.9\% $\pm$ 0.6, rand score: 0.78 $\pm$ 0.03 vs. 0.71 $\pm$ 0.05). It is interesting to note that the number of parameters in the contacting path is almost half the number of parameters in the expanding path, but still we get better results by training fewer number of parameters. Figure \ref{fig3} represents some examples of the results on the test set. Contrary to ultrasound images, chest X-ray images resulted in an almost equal evaluation scores when segmenting lungs, by applying the same fine-tuning procedure (Dice score: 0.98 for both scenarios, pixel error: 1.1\% vs. 1.3\%, rand score: 0.95 for both scenarios).
\\
When we investigated fine-tuning in a more rigorous manner (including blocks of layers one-by-one), better performance was achieved moving from shallow to deep layers compared to moving from deep to shallow layers. Figure \ref{fig4} shows the average Dice score for the two directions of fine-tuning. Note that the middle part of Figure \ref{fig4} corresponds to the results reported in the previous paragraph; blue graph: fine-tuning the contracting path and freezing the expanding path, red graph: fine-tuning the expanding path and freezing the contracting path.  

\begin{figure}
\includegraphics[width=\textwidth]{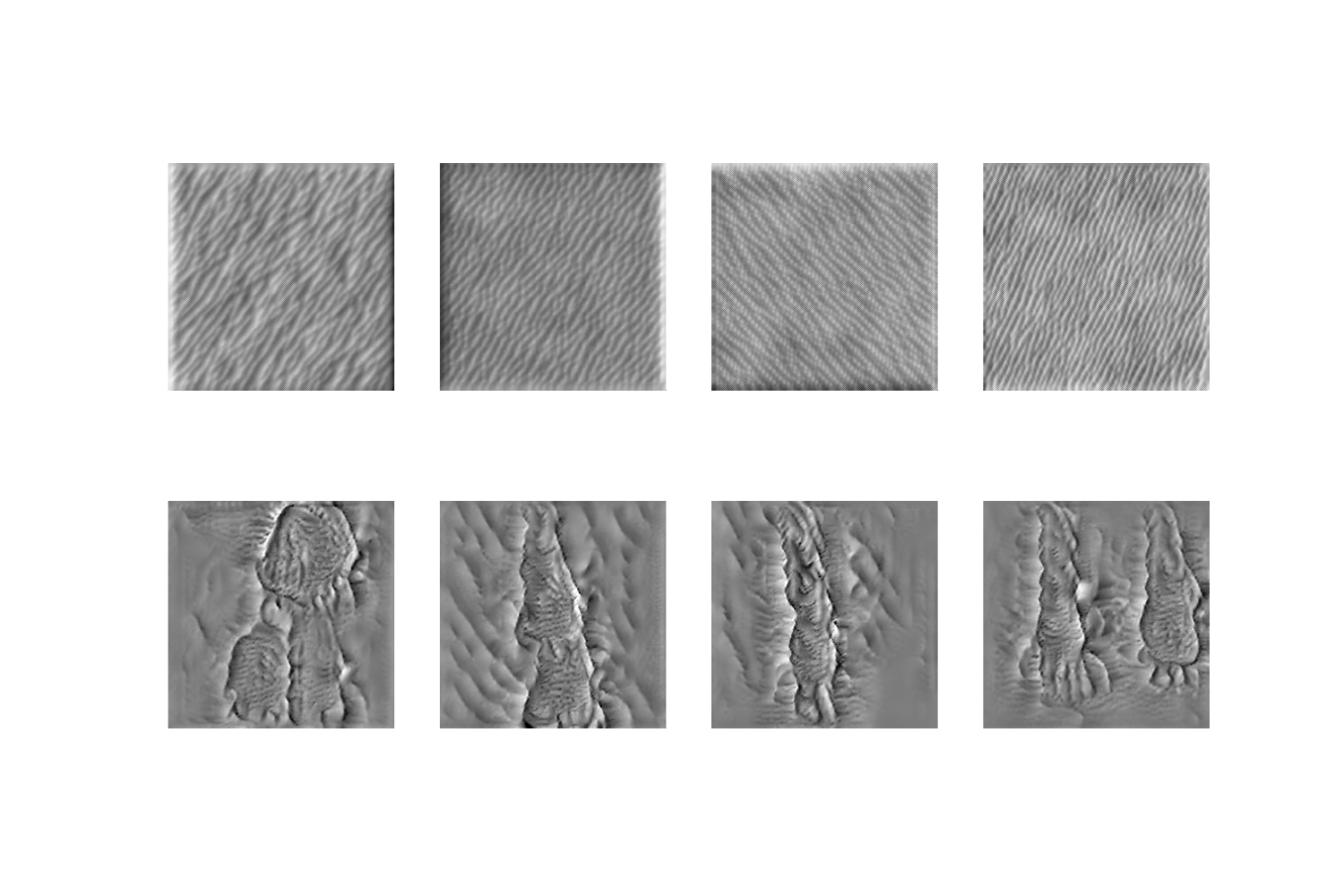}
\vspace{-1.5cm}
\caption{Some examples of images which maximize the activation in a network trained on ultrasound images, top: in the fourth layer, bottom: in the deepest convolutional layer.} \label{fig5}
\end{figure}

As the number of parameters in the expanding part of the network is much higher than the contracting part, it would be expected that the expanding part of the network trains more slowly than the contracting part, and it would therefore affect the results. We used a fixed number of epochs (20 epochs) for fine-tuning the network in all studied scenarios. Although the network was stable after 20 epochs, we added 20 more epochs (40 epochs in total now) to examine the impact of number of epochs on the segmentation performance. The changes in Dice score were below 1\% for all cases except for the case when the deepest block of layers was trained and the rest of the network was fixed (first point in figure \ref{fig4}); the Dice score improved by 3\%, but it was still much lower than the other path. Adding 10 more epochs did not change the results anymore. Thus, the effect of the speed of training is not a major one.
\\
In order to see what features are seen by different layers of the network, we employed Keras-Vis \cite{Kotikalapudi2017} to visualize the input image which maximize the activation in each neuron. Several low-level patterns were recognized for shallow layers (mostly edges), while high-level maps are more detailed and complex shapes. Figure \ref{fig5} shows some examples in different neurons of shallow and deep layers. 
\section{Discussion and Conclusions}
We showed that in breast ultrasound image segmentation using U-Net, fine-tuning shallow layers of a pre-trained network outperforms fine-tuning deep layers, when a small number of images are available. It could be due to the presence of specific low-level patterns such as speckles in this modality, which are associated with shallow layers of the network.  

It is important to note that U-Net is not a simple feedforward architecture. The notion of deep and shallow is ambiguous in a U-Net, because there are short and long paths from the input to the output. In this study, we considered the depth of a layer to be the longest possible path to reach it. Some differences in the behavior could potentially be related to the difference in architecture of a U-Net. However, given that the behaviour on non-ultrasound data is similar to previously reported results, we believe the primary cause of the differences are due to the character of the image.

\section*{Acknowledgment}
This work was supported by in part by Natural Science and Engineering Research Council of Canada (NSERC) Discovery Grant RGPIN-2015-04136.

%
%
%
\bibliographystyle{splncs04}
\bibliography{mybibliography}

\end{document}